\documentclass[12pt]{iopart}

\usepackage{graphicx}
\usepackage{bm}
\usepackage{amsfonts}
\usepackage{color}
\usepackage[up]{subfigure}

\usepackage{latexsym}
\usepackage{amssymb}
\usepackage[colorlinks=true, citecolor=blue, urlcolor=blue]{hyperref}
\usepackage{float}

\usepackage{amsfonts}
\usepackage{bigfoot}
\DeclareNewFootnote[para]{B}[alph]
\DeclareNewFootnote{default}

\usepackage{perpage}
\MakePerPage{footnoteB}

\begin{document}

\title[]{The Hardy's nonlocality argument}

\author{Sujit K Choudhary $^1$ and Pankaj Agrawal$^1$}
\address {$^1$ Institute of Physics, Sachivalaya Marg,
Bhubaneswar-751005, Odisha, India.}
\ead{sujit@iopb.res.in, agrawal@iopb.res.in}

\begin{abstract}
Certain predictions of quantum theory are not compatible with the notion of local-realism.
This was the content of Bell's famous theorem of the year 1964. Bell proved this with the help of an
inequality, famously known as Bell's inequality. The alternative proofs of Bell's theorem 
without using Bell's inequality are known as `nonlocality without inequality (NLWI)' proofs.
We, review one such proof, namely the Hardy's proof which due to its simplicity
and generality has been considered the best version of Bell's theorem. 

\end{abstract}
\vspace{2pc}

\section{Introduction}

Many of the current methods and developments in quantum
information processing have grown out of a long struggle of
physicists with the foundations of quantum theory. One of the most
significant example is the famous consideration by
Einstein, Podolsky and Rosen \cite{EPR} on reality, locality and
completeness of a physical theory. The standard Quantum Theory 
is essentially a statistical theory. It gives accurate predictions for statistical 
distributions of outcomes obtained in a real experiment. However, it does not tell which 
outcome will be observed in a particular measurement-experiment unless the state undergoing the measurement
is an eigenstate of the observable being measured. Interestingly, it also does not disallow for 
a finer theory where the outcome of an individual measurement  may be determined
by some variables outside the domain of definition of quantum theory.
The statistical distributions of quantum theory would then be averages over these
hidden variables. Such Hidden variable models indeed exist \cite{bellrev}. However, as shown by Bell in 1964 \cite{bell}, there exist correlations in nature, to explain which, such models are bound to be nonlocal 
(in the sense that such models must have a mechanism whereby the
setting of one measuring device can influence the reading of another device howsoever remote). This incompatibility
of quantum phenomena with the dual assumptions of locality and reality (determinism) is known 
as Bell's theorem. Bell proved this theorem by means of an inequality; famously known as Bell's inequality.
But, Bell's inequality is not the only way to prove Bell's theorem. The alternative proofs of Bell's theorem
without using Bell's inequalities are called `nonlocality without inequality (NLWI)  proofs'.  
Unlike the case of Bell's inequality where we collect statistics of many events, in these proofs the focus is on a single event whose occurrence shows the incompatibility of quantum theory with the notion of local-realism.  
The first such proof is due to Greenberger, Horne and Zeilinger \cite{green}. 
In their argument, they used correlations of a state of four
spin- 1/2 particles: $\frac{1}{\sqrt{2}}[|0000\rangle-|1111\rangle]$
and remarked that this argument also holds for the three-qubit analog of the above state, namely for $\frac{1}{\sqrt{2}}[|000\rangle-|111\rangle]$. Although their proof is direct, it requires at least an eight-dimensional Hilbert space. In 1992, Hardy \cite{hardy92} gave a proof of Bell's theorem (without inequality)
which like Bell's proof, requires only two qubits. The Hardy's
argument of ``nonlocality without inequiality" has been considered to be the
``best version of Bell's theorem"\cite{mermin94}. 

Recent years have witnessed considerable amount of interest in Hardy's and Hardy-like
nonlocality arguments. During these years, this argument has been generalized for 
various systems ranging from two qudits, multiqubits and to systems exhibiting temporal nonlocality. 
On one hand, it has been found to be useful in understanding the relation between quantum theory and special relativity, while on the other it has found applications in  
many information processing tasks like cryptography, randomness certification, dimension witnessing. We briefly review all these developments. Organization of the paper is as follows. In Sec.2, we briefly review    
the ontological framework of an operational theory as this will subsequently be used in Hardy's nonlocality
argument. Section 3 presents the Hardy's nonlocality argument. Section 4 deals with the two-qubit states exhibiting Hardy's nonlocality followed by extension of Hardy's argument for two-qudit systems in Sec. 5. Hardy's nonlocality in the framework of generalized nonlocal theory has been described in Sec. 6; Sec. 7 deals with the various generalizations of this argument in mutipartite scenarios. In Sec. 8, we describe the role of Hardy's nonlocality in some information theoretical tasks. Section 9 deals with Hardy-like argument for temporal nonlocality and Sec. 10 concludes.    

\section{Ontological model of an operational theory}
We, in the following, briefly describe the ontological framework of an operational theory (for details of this framework, we refer to \cite{spek05,rudolph}), as we will use it subsequently in the Hardy's nonlocality argument.   

Quantum Theory is example of an  operational theory. The goal  of an operational theory is merely to specify the probabilities $p(k|M,P,T)$ of different outcomes $k\in\mathcal{K}_M$ that may result from a measurement procedure $M\in\mathcal{M}$ given a particular preparation procedure $P\in\mathcal{P}$, and a particular transformation procedure $T\in\mathcal{T}$; where $\mathcal{M}$, $\mathcal{P}$ and $\mathcal{T}$ respectively denote the sets of measurement procedures, preparation procedures and transformation procedures; $\mathcal{K}_M$ denotes the set of measurement results for the measurement M. 

Whereas an operational theory does not tell anything about \emph{physical state} of the system, in an ontological model of an operational theory, the primitives of description are the actual state of affairs of the system. A preparation procedure is assumed to prepare a system with certain properties and a measurement procedure is assumed to reveal something about those properties. A complete specification of the properties of a system is referred to as the ontic state of that system. In an ontological model for quantum theory, a particular preparation method $P_{\psi}$ which prepares the quantum state $|\psi\rangle$, actually puts the system into some ontic state $\lambda\in\Lambda$, $\Lambda$ denotes the ontic state space. An observer who knows the preparation $P_{\psi}$ may nonetheless have incomplete knowledge of $\lambda$. Thus, in general, an ontological model associates a probability distribution $p(\lambda|P_{\psi})$ with preparation $P_{\psi}$ of $|\psi\rangle$. $p(\lambda| P_{\psi})$ is called the \emph{epistemic state} as it encodes observer's \emph{epistemic ignorance} about the state of the system. It must satisfy
\begin{equation}\nonumber
\int_{\Lambda}p(\lambda|P_{\psi})d\lambda=1~~
\forall~|\psi\rangle~\mbox{and}~P_{\psi}.
\end{equation}
Similarly, the model may be such that the ontic state $\lambda$ determines only the probability $\xi(k|\lambda,M)$, of different outcomes $k$ for the measurement method $M$. However, in a deterministic model  $\xi(k|\lambda,M)\in \{0,1\}$. The response functions $\xi(k|\lambda,M)\in[0,1]$, should satisfy
\begin{equation}\nonumber
\sum_{k\in\mathcal{K}_M}\xi(k|\lambda,M)=1~~\forall~~\lambda,~~M.
\end{equation}
Thus, in the ontological model, the probability $p(k|M,P)$ is specified as
\begin{equation}\nonumber
p(k|M,P)=\int_{\Lambda} \xi(k|M,\lambda)p(\lambda|P) d\lambda.
\end{equation} 
As the model is required to reproduce the observed probabilities (quantum predictions) hence the following must also be satisfied
\begin{equation}\nonumber
\int_{\Lambda} \xi(\phi|M,\lambda)p(\lambda|P_{\psi}) d\lambda = |\langle\phi|\psi\rangle|^2.
\end{equation} 

The transformation processes $T$ are represented by stochastic maps from ontic states to ontic states. $\mathcal{T}(\lambda'|\lambda)$ represents the probability distribution over subsequent ontic states given that the earlier ontic state one started with was $\lambda$.

\subsection{The Hardy's scenario}
A typical Hardy's experiment involves two spatially separated observers, Alice
and Bob, who share a physical system consisting of two subsystems. 
They can perform measurements on the subsystems in their possessions and 
collect statistics to calculate the joint probabilities $p(a,b|A, B,P)$. Here $A$ and $B$ 
denote the observables chosen respectively by Alice and Bob; $a$ and $b$ are the corresponding outcomes. 
Each pair of subsystems is prepared by an agreed-upon reproducible procedure $P$ (which in quantum theory is represented by quantum state for the pair of subsystems). 

As described earlier, an ontological model for this
experiment  consists of the ontic variables $\lambda$ belonging to the ontic state space $\Lambda$, a probability
distribution $p(\lambda| P)$ for the preparation procedure $P$ and the conditional probability 
$p(a,b|A,B, P,\lambda)$. The prediction for the observed joint probability by this model (given in the LHS of the equation below) must match with the observed probability, i.e., 
\begin{equation}\label{reproducibility}
\int_{\Lambda} p(a,b|A,B, P,\lambda)p(\lambda| P)~d\lambda={\rm Prob}(a,b|A,B, P).
\end{equation}
 The Integrand of the above equation, by Baye's theorem, can be rewritten as
\begin{equation}\label{bays}
  p(a,b|A,B, P,\lambda)=p(a|A,B, P,\lambda)~p(b|A,a,B, P,\lambda).
\end{equation}
An ontological model is said to be deterministic iff $ p(a,b|A,B, P,\lambda)\in\{0,1\}$ $ \forall a, b, A, B$ \cite{wisemanjphys}.
This implies that the outcome of Bob's measurement (and similarly also of Alice's measurement) is determined 
by $A$, $B$, $P$ and $\lambda$ only. Thus, for a deterministic ontological model
\begin{equation} \label{determinism}
 p(b|A,a,B, P,\lambda)=p(b|A,B, P,\lambda).
\end{equation}
A model is said to satisfy locality iff 
\begin{eqnarray}\label{nonlocality}
 p(a|A,B, P,\lambda)=p(a|A, P,\lambda)~\forall~ a, A, B,\nonumber\\
 p(b|A,B, P,\lambda)=p(b|B, P,\lambda)~\forall~ b, A, B.
\end{eqnarray}
Using Eqs. (\ref{determinism}) and (\ref{nonlocality}) in Eq. (\ref{bays}), we get
\begin{equation}\label{factorizability}
  p(a,b|A,B, P,\lambda)=p(a|A, P,\lambda)p(b|B, P,\lambda).
\end{equation}
Thus in a local-deterministic model, Eq. (\ref{reproducibility}), takes the following form 
\begin{equation}\label{factorizability1}
 {\rm Prob}(a,b|A,B, P)=\int_{\Lambda} p(a|A, P,\lambda)~p(b|B, P,\lambda)p(\lambda| P)~d\lambda.
\end{equation}

\section{The Hardy's nonlocality argument}

Consider a physical system consisting of two subsystems shared
between Alice and Bob. The two observers (Alice and Bob) have
access to one subsystem each. For each pair of subsystems,
the choices of observables and their respective outcomes occur 
in regions which are space-like separated from each other. Assume that Alice can run the
experiments of measuring any one (chosen freely) of the two
$\{+ 1, - 1\}$-valued random variables $A$  and $A^{'}$
corresponding to her subsystem whereas Bob can run the experiments
of measuring any one (chosen freely) of the two $\{+ 1, -
1\}$-valued random variables $B$  and $B^{'}$ corresponding to the
subsystem in his possession. The two subsystems undergoing measurement are prepared by an agreed-upon reproducible procedure $P$. 

Consider now the following four conditions:
\begin{eqnarray}
{\rm Prob} (+1,+1|A,~B,~ P) &>& 0,\label{hardy2q1}\\
{\rm Prob} (-1, +1|A^{'},B, ~P) &=& 0,\label{hardy2q2}\\
{\rm Prob} (+1,-1|A,~ B^{'}, ~P) &=& 0,\label{hardy2q3}\\
{\rm Prob} (+1,+1|A^{'},~ B^{'},~P) &=& 0.\label{hardy2q4}
\end{eqnarray}

The above four conditions together form the basis of Hardy's
nonlocality argument. The first condition says that in an experiment in which
Alice chooses to measure the observable $A$ and Bob chooses the observable
$B$, the probability that both will get $+1$ as measurement outcomes is nonzero.
Other conditions can be analyzed similarly. The Hardy's nonlocality argument 
makes use of the fact that these four conditions cannot be fulfilled
simultaneously in the framework of a local-realistic theory, 
but they can be in quantum mechanics. To see this, we start with Eq.(\ref{hardy2q1}). By using Eq.(\ref{factorizability1}) in the first Hardy condition (Ineq.(\ref{hardy2q1})), we notice that in a local-realistic  theory, this condition will get satisfied if
\begin{equation}\label{condition1}
 \int_{\Lambda} p(+1|A, P,\lambda)~p(+1|B, P,\lambda)p(\lambda| P)~d\lambda >0.
\end{equation}
This implies the existence of a subset $\Lambda'$ of the ontic state space $\Lambda$ in which
$p(+1|A, P,\lambda)>0$ and $p(+1|B, P,\lambda)>0$. Use of Eq.(\ref{factorizability1}) in Eq.(\ref{hardy2q2})
says that for the local-realistic model to satisfy the second Hardy's condition, $p(-1|A^{'}, P,\lambda)~p(+1|B, P,\lambda)=0$ for all $\lambda\in\Lambda$. However, for $\lambda\in \Lambda'$, 
this condition implies $p(-1|A^{'}, P,\lambda)=0$ or equivalently for these $\lambda$'s,  
$p(+1|A^{'}, P,\lambda)=1$. Similar reasoning for the third Hardy's condition provides $p(+1|B^{'}, P,\lambda)=1$
for $\lambda\in \Lambda'$. Thus a local-realistic model will predict for the last probability in the Hardy's condition (Eq.\ref{hardy2q4}) as:
\begin{eqnarray}\label{contradiction}
 {\rm Prob} (+1,+1|A^{'},~ B^{'},~P) &=\int_{\Lambda} p(+1|A^{'}, P,\lambda)~p(+1| B^{'}, P,\lambda)p(\lambda P)~d\lambda \nonumber\\
&\geq\int_{\Lambda'} p(+1|A^{'}, P,\lambda)~p(+1| B^{'}, P,\lambda)p(\lambda| P)~d\lambda\nonumber\\
&=\int_{\Lambda'} p(\lambda| P)~d\lambda >0.
\end{eqnarray}

However, there are quantum states which satisfy these conditions. 
In fact, almost all pure entangled states of two-qubits satisfy these conditions (maximally entangled states 
are the exceptions) \cite{hardy93, Goldstein, jordan}. The nonzero probability  appearing in the argument (say q) is called the success probability of this argument as this is equal to the fraction of runs in which quantum predictions contradicts the reasoning based on local-realism. 

\section{Hardy nonlocality for two-qubits}
\subsection{Every pure nonmaximally entangled state of two-qubits exhibits Hardy's nonlocality}

To see this, we first notice that   
any pure nonmaximally entangled state $| \psi
\rangle$\footnote{A pure entangled state $| \psi \rangle$ of two-qubits
is said to be maximally entangled if the
reduced density matrices corresponding to each particle is
proportional to the identity matrix $I$ in two dimension.} of two
spin-1/2 particles can be written as \cite{Goldstein}
\begin{equation}\label{goldsteinstate}
\label{eqn24} | \psi \rangle = a| v_1 \rangle \otimes | v_2
\rangle + b| u_1 \rangle \otimes | v_2 \rangle + c | v_1 \rangle
\otimes | u_2 \rangle  \,(abc \ne 0)
\end{equation}
for a proper choice of orthonormal basis $\left\{| u_i \rangle, |
v_i \rangle\right\}$ for $i$-th particle, $i = 1, 2$;  $| u_1 \rangle, |
u_2 \rangle$ need not bear any relationship with each other. 

It can easily be checked that the above state (\ref{goldsteinstate}) 
will satisfy all the four Hardy's conditions (\ref{hardy2q1})-(\ref{hardy2q4}) for the following choice of observables:  
\begin{eqnarray}\label{goldsteinobservables} 
A = | w_1^\perp \rangle \langle  w_1^\perp| -| w_1 \rangle \langle w_1|,\nonumber\\
A'= | u_1 \rangle \langle u_1 |-| v_1 \rangle \langle v_1|,\nonumber\\
 B=| w_2^\perp \rangle \langle  w_2^\perp| -| w_2 \rangle \langle w_2|,\nonumber\\
B'= | u_2 \rangle \langle u_2 |-| v_2 \rangle \langle v_2|
\end{eqnarray}
where
\begin{eqnarray}\label{goldsteinbasis}
| w_1 \rangle = \frac{a| v_1 \rangle + b| u_1 \rangle} {\sqrt{|a|^2 + |b|^2}},\nonumber\\
| w_2 \rangle = \frac{a| v_2 \rangle + c| u_2 \rangle} {\sqrt{|a|^2 + |c|^2}}.
\end{eqnarray}

\subsection{Maximally entangled state and product states of two-qubits do not exhibit Hardy's nonlocality }

A maximally entangled state of two-qubits shared between two parties Alice and Bob can be written as $| \phi
\rangle_{AB} = \frac{1}{\sqrt{2}} \left(| u \rangle_A \otimes |u
\rangle_B + | v \rangle_A \otimes |v\rangle_B\right)$ (where $\left\{| u_i \rangle, |
v_i \rangle\right\}$ are the orthonormal bases for $i$-th particle, $i = A, B$). 
The state $| \phi\rangle_{AB}$ will be said to exhibit Hardy's nonlocality, if there exists
a set of observables $A$ and $A'$ for Alice and a set $B$ and $B'$ for Bob, such that all the 
four Hardy's conditions (\ref{hardy2q1})-(\ref{hardy2q4}) are satisfied simultaneously.  
The most general observables on Alice's and Bob's side can be written as
\begin{eqnarray}\label{maximallyentangledobservables}
A = | w_1 \rangle \langle w_1 |-| w_1^{\perp} \rangle \langle  w_1^{\perp}|,\nonumber\\
A'= | w'_1 \rangle \langle w'_1 |-|{ w'_1}^{\perp }\rangle \langle  {w'_1}^{\perp}|,\nonumber\\
B = | w_2 \rangle \langle w_2 |-| w_2^{\perp} \rangle \langle  w_2^{\perp}|,\nonumber\\
B'= | w'_2 \rangle \langle w'_2 |-|{ w'_2}^{\perp }\rangle \langle  {w'_2}^{\perp}|\nonumber\\
\end{eqnarray} 
where
\begin{eqnarray}\label{maximallyentangledbasis}
| w_1 \rangle = \frac{a| u \rangle_A + b| v \rangle_A} {\sqrt{|a|^2 + |b|^2}},\nonumber\\
| w'_1 \rangle = \frac{c| u \rangle_A + d| v \rangle_A} {\sqrt{|c|^2 + |d|^2}},\nonumber\\.
| w_2 \rangle = \frac{e| u \rangle_B + f| v \rangle_B} {\sqrt{|e|^2 + |f|^2}},\nonumber\\
| w'_1 \rangle = \frac{g| u \rangle_B + h| v \rangle_B} {\sqrt{|g|^2 + |h|^2}}
\end{eqnarray}
and $ \langle w_1 |w_1^{\perp} \rangle=\langle w_2 |w_2^{\perp} \rangle= \langle w'_1 |{ w'_1}^{\perp }\rangle=   \langle w'_2 |{ w'_2}^{\perp }\rangle =0$.
For the satisfaction of the  last three  Hardy's condition [Eq. (\ref{hardy2q2})-Eq.(\ref{hardy2q4} )] , the following should hold
\begin{eqnarray}
\langle \phi_{AB}| { w'_1}^{\perp }w_2\rangle=0~\Rightarrow ed^{*}=c^{*}f,\label{maximallyentangledcondition1}\\
\langle \phi_{AB}| w_1{ w'_2}^{\perp }\rangle=0\Rightarrow ah^{*}=bg^{*},\label{maximallyentangledcondition2}\\
\langle \phi_{AB}|  w'_1 w'_2\rangle=0~\Rightarrow cg=-hd, \label{maximallyentangledcondition3}
\end{eqnarray}
where $|  w'_1 w'_2\rangle$  represents $|  w'_1\rangle\otimes |w'_2\rangle$ and so on ; $h^{*} g^{*}, c^{*}, d^{*}$  denote the complex conjugates of $h,g,c$ and $d$ respectively.
By multiplying Eqs. (\ref{maximallyentangledcondition1}) and (\ref{maximallyentangledcondition1}) and then using Eq. (\ref{maximallyentangledcondition3}), we get
\begin{equation}\nonumber
 c^{*} g^{*}(bf+ae)=0.
\end{equation}
As the observables $A$ and $A'$ are noncommuting; B and B' also do not commute , hence $c^{*} g^{*}\neq0$ and so $bf+ae=0$ . But, this renders  the last Hardy probability (\ref{hardy2q1}), ${\rm Prob} (+1,+1|A,~B,~ P) =|\langle \phi_{AB}| w_1w_2\rangle|^2=|bf+ae   |^2$ equal to zero. Thus, no maximally entangled state of two-qubits can satisfy all the Hardy's conditions.  A similar logic can be applied to show that product states also do not exhibit this nonlocality.

\subsection{No mixed entangled state of two-qubits exhibits Hardy's nonlocality }
We have seen in earlier paragraphs that almost all pure entangled state of two-qubits can exhibit Hardy's nonlocality
for suitably chosen observables. The choice of such a set of observables is not unique for a given entangled state. 
This was shown by Jorden in \cite{jordan}. He showed that for a particular entangled state, there are many choices of set of observables which satisfy Hardy's nonlocality conditions. In \cite{jordan}, it has also been shown that for any choice of two different measurement possibilities for each particle, a state can be found that satisfies
all the Hardy's conditions. Interestingly, this state was shown to be unique for a system of two-qubits \cite{Kar(1997a)}. In \cite{Kar(1997a)}, Kar showed that for a system of two spin-1/2 particles and for four arbitrary spin observables, two for each and noncommuting \footnote{It can easily be noticed that for running the Hardy's argument,
the two observables on Alice's side should not commute and similarly similarly the observables on Bob's side should also be noncommuting}, there exists a unique state satisfying all the Hardy's condition, i.e., no mixed state of such a system can exhibit Hardy's nonlocality \footnote{This is in contrast with Bell's nonlocality which some mixed state of two-qubits exhibit by violating it.}  

To see this we write the observables involved in Hardy's argument Eqs.(\ref{hardy2q1})-(\ref{hardy2q4}) as follows
\begin{eqnarray}
A =  |a_1 \rangle \langle a_1 |-| a_1^\perp \rangle \langle  a_1^\perp|,\nonumber\\ 
A'= |a'_1 \rangle \langle a'_1 |-| {a'_{1}}{^\perp} \rangle \langle  {a'_{1}}^{\perp}|,\nonumber\\
B =  |b_1 \rangle \langle b_1 |-| b_1^\perp \rangle \langle  b_1^\perp|,\nonumber\\ 
B'= |b'_1 \rangle \langle b'_1 |-| {b'_{1}}{^\perp} \rangle \langle  {b'_{1}}^{\perp}|.
\end{eqnarray}
A state vector which satisfy all the Hardy's condition must be orthogonal to 
 the vectors $|\phi_1 \rangle=|{a'_{1}}{^\perp} \rangle\otimes|b_1 \rangle$, $|\phi_2 \rangle=|a_1 \rangle\otimes| {b'_{1}}{^\perp} \rangle$ and  
$|\phi_3 \rangle=|a'_1 \rangle\otimes|b'_1 \rangle$ and nonorthogonal to $|\phi_4 \rangle=|a_1 \rangle\otimes |b_1 \rangle$.
As  $A$ does not commute with $A'$ and  $B$ and $B'$ are also noncommuting, hence the above four vectors are linearly
independent and spans the four dimensional Hilbert space associated with the system. The vector which is orthogonal to the three-dimensional subspace generated by the first three vectors is unique therefore and this vector is also nonorthogonal to the last vector. Hence, we conclude that for any choice of a set of four spin observables of above mentioned type, there exists only one state satisfying the Hardy's condition and therefore no mixed state of two spin-1/2 particles exhibit Hardy's nonlocality.

\subsection{The maximum Hardy's probability}
We have seen in the previous paragraphs that in case of  two-qubits systems only nonmaximally entangled states exhibits Hardy's nonlocality for suitably chosen observables. The canonical form
of such states and the set of observables for which they can exhibit Hardy's nonlocality are given 
in section (4.1). From  equations (\ref{goldsteinstate}), (\ref{goldsteinobservables}) and 
(\ref{goldsteinbasis}), the nonzero Hardy's probability (\ref{hardy2q1})  becomes
\begin{eqnarray}\label{hardymaximum}
{\rm Prob} (+1,+1|A,~B) &=|\langle \psi| w_1^\perp\rangle\otimes |w_2^\perp \rangle |^2\nonumber\\
&=\frac{|a|^2 |b|^2|c|^2}{(|a|^2 + |b|^2)( |a|^2 + |c|^2)}.
\end{eqnarray}
The maximum of this function can easily be calculated by using the relation $|a|^2 +|b|^2|+|c|^2=1$  
  and noting that $abc\neq0$. The maximum comes out to be equal to  $\frac{5\sqrt{5}-11}{2}$
for $|b|=|c|=\sqrt{\frac{3-\sqrt{5}}{2}}$.

\section{Hardy's nonlocality for two-qudits}

The Hardy's conditions for bipartite higher dimensional systems were first generalized 
 by Clifton and Niemann \cite{clifton92} which was later put in its minimal form by Kunkri and Choudhary in \cite{kunkri}.  The minimal form has been further studied in \cite{ghosh2011,scarani}. The maximum probability of nonlocal events has been found to remain  the same irrespective of the dimension of local subsystems.  Recently, Cabello \cite{cabello2013} has introduced another generalization of Hardy's conditions. The probability of nonlocal events for two-qudit systems grows with the dimension of the local systems in this generalization. In the following, we briefly review all these developments.

\subsection{The generalization of Cliffton and Niemann }

Hardy's argument was generalized for two spin--$s$($s =
\frac{1}{2}, 1, \frac{3}{2}.......$) particles by Clifton and
Niemann\cite{clifton92}. Consider two spin-$s$ particles $A$ and $B$ which are far
separated from each other and possessed respectively by two observers Alice and Bob. Let $S_a$ and $S_{a^{\prime}}$
represent spin component of particle $A$ along the directions $\hat{a}$ and
$\hat{a^{\prime}}$ respectively. Similarly $S_b$ and $S_{b^{\prime}}$
represent spin component of particle $B$ along the directions $\hat{b}$
and $\hat{b^{\prime}}$. The values of each of the components $S_a$, $S_b$, $S_{a^{\prime}}$,
$S_{b^{\prime}}$ runs from $-s$ to $+s$ in step of one.
Consider now the following set of equations:
\begin{equation}
\label{cliffton} \left.
\begin{array}{lcl}
{\rm Prob} (S_a = S_b = s) = 0,\\
{\rm Prob} (S_a + S_{b^{\prime}} \ge 0) = 1,\\
{\rm Prob} (S_{a^{\prime}} + S_b \ge 0) = 1,\\
{\rm Prob} (S_{a^{\prime}}= S_{b^{\prime}} = -s) = q,
\end{array}
\right\}~ ({\rm with}~ q > 0).
\end{equation}
 It can be checked that the set of above four equations are incompatible with the notion of local-realism. The last equation of the above set implies
that (i) there is a non-zero probability (which is $q$ here) of simultaneous occurrence of $S_{a^{\prime}} =
-s$ and $S_{b^{\prime}} = -s$. This will then imply that (according to
classical probability theory) (ii) if Alice chooses to perform the
measurement of $S_{a^{\prime}}$, there will be a non-zero probability (which
should be at least $q$) of getting the value $-s$, and similarly,
(iii) if Bob chooses to perform the measurement of $S_{b^{\prime}}$, there will
be a non-zero probability (which should be at least $q$) of
getting the value $-s$. Now the third condition in equation
(\ref{cliffton}) implies that if Alice chooses for the measurement of
$S_{a^{\prime}}$ and Bob chooses for $S_{b}$, he is bound to get the value
$s$ whenever Alice gets the value $-s$ (which Alice can indeed get
with a non-zero probability, according to (ii)). Similarly the
2nd condition in equation (\ref{cliffton}) implies that if Alice
chooses to measure $S_{a}$ and Bob chooses to perform
$S_{b^{\prime}}$, she is bound to get the value $s$ whenever Bob gets
the value $-s$ (which Bob can indeed get with a non-zero
probability, according to (ii)). So $S_{a}=s$ is a `reality'
of $S_{a}$ while $S_{b}=s$ is also a `reality' of $S_{b}$,
according to EPR\cite{EPR}. Now, invoking `locality', $S_{a}=s$ and $S_{b}=s$ is a joint `reality' of the composite system. And, according to condition (i) (again, using classical
probability arguments), the joint probability of occurrence of
$S_a = S_b = s$ must be at least $q$. This contradicts
the first condition of equation (\ref{cliffton}). However, 
as shown in \cite{clifton92}, for a given choice of
observables, there is always a quantum state which satisfies the above set of equations. Later, using the
above argument, Ghosh and Kar \cite{ghosh98} showed that for any
two spin-$s$ ($s = 1, \frac{3}{2}, 2.......$) particles and for
two measurement possibilities for each of them, there are
infinitely many states exhibiting this type of nonlocality. Hence
there are mixed states of two spin-$s$ ($s \ge 1$) particles which
would exhibit this type of nonlocality. This is in contrast to the qubit case
where no two-qubit mixed state exhibits Hardy's nonlocality.

\subsection{Hardy's conditions in minimal form}
A closer look of the nonlocality conditions given above reveals that these conditions include
some equations which play no role in establishing the nonlocality for
quantum states. To make this more explicit, we
first consider a system of two spin-1 particles. For such a system
the above equations can be written as

\begin{equation}\label{clifton1}
{\rm Prob}(S_a = +1, S_b = +1) = 0,
\end{equation}

\begin{equation}\label{clifton2}
{\rm Prob}(S_a = -1, S_{b^{\prime}} = -1) = 0,
\end{equation}

\begin{equation}\label{clifton3}
{\rm Prob}(S_a = -1, S_{b^{\prime}} = 0 ) = 0,
\end{equation}

\begin{equation}\label{clifton4}
{\rm Prob}(S_a = 0, S_{b^{\prime}} = -1) = 0,
\end{equation}

\begin{equation}\label{clifton5}
{\rm Prob}(S_{a^{\prime}} = -1, S_b = -1) = 0,
\end{equation}

\begin{equation}\label{clifton6}
{\rm Prob}(S_{a^{\prime}} = -1, S_b = 0) = 0,
\end{equation}

\begin{equation}\label{clifton7}
{\rm Prob}(S_{a^{\prime}} = 0, S_b = -1) = 0,
\end{equation}

\begin{equation}\label{clifton8}
{\rm Prob}(S_{a^{\prime}} = -1, S_{b^{\prime}} = -1) = q.
\end{equation}

For showing the incompatibility of the above equations with the
notion of local-realism, we start with equation (\ref{clifton8}). This
equation tells that if there is an underlying local-realistic
theory then there are some ontic states for which
$S_{a^{\prime}} = -1$, $S_{b^{\prime}} = -1$. Now, for these states, 
equations (\ref{clifton2}) and (\ref{clifton4}) tell us that $S_a = +1$. Similarly
for these states, $S_b = +1$ according to equations(\ref{clifton5}) and (\ref{clifton6}). So, ${\rm Prob}(S_a = +1, S_b = +1)$ should have been
non-zero but this contradicts  equation(\ref{clifton1}). 

It is noteworthy here that in order to run the nonlocality argument
for $s = 1$, we have not used equations (\ref{clifton3}) and  (\ref{clifton7}) . So
these equations are redundant and are unnecessarily restricting
the set of states exhibiting this type of  non-locality.\footnote{ The
states exhibiting the non-locality need not be orthogonal to the
projectors $P[|S_a = -1, S_{b^{\prime}} = 0 \rangle ]$ and
$P[|S_{a^{\prime}} = 0, S_b = -1\rangle ]$ appearing in
(\ref{clifton3}) and (\ref{clifton7}). These need to be  orthogonal only to the
projectors $P[|S_a = +1, S_b = +1\rangle ]$, $P[|S_a = -1,
S_{b^{\prime}} = -1,\rangle ]$, $P[|S_a = 0, S_{b^{\prime}} =
-1\rangle ]$,$P[|S_{a^{\prime}} = -1, S_b = -1 \rangle ]$and
$P[|S_{a^{\prime}} = -1, S_b = 0
\rangle ]$.}

These restrictions have been taken care of in \cite{kunkri}, where for a system of 
two $d$ level particles, the Hardy's conditions have been generalized as follows:
\begin{equation}\label{minimalform}
\begin{array}{lcl}
{\rm Prob}(A_1,a_1~;\ B_1,b_1)=0, \label{eqn-1}\\
{\rm Prob}(A_1,\neg a_1~;\ B_2,b_2)=0,\label{eqn-2}  \\
{\rm Prob}(A_2,a_2~;\ B_1,\neg b_1)=0, \label{eqn-3}\\
 {\rm Prob}(A_2,a_2~;\ B_2,b_2) \ =p (>0) \label{eqn-4}.
\end{array}
\end{equation}
It can be easily seen \cite{kunkri} that in a situation where a two $d$-level physical system 
is shared between two far separated observers Alice and Bob having access to one subsystem each
and where the observers Alice and Bob can choose to measure one of the two observables $A_1$ or $A_2$ and  $B_1$ or $B_2$ on their relative subsystems, the above four conditions \footnote{The first condition says that if Alice chooses to measure the observable $A_1$ and 
Bob  chooses observable $B_1$, he will not obtain  $b_1$ as measurement result
whenever Alice has  detected the measurement value $a_1$. The remaining equations can be analyzed
in a similar manner ($\neg a_i$ denotes a measurement with any result other than $a_i$
and similarly $\neg b_j$ denotes a measurement with any result other than $b_j$)} cannot be satisfied simultaneously by a local-realistic reasoning. However, they can be in quantum mechanics. As usual, the nonzero probability $p$ appearing above is the success probability of this argument in showing nonlocal features of quantum states.

\subsection{Maximum probability of nonlocal events}
This form of Hardy's argument has been further studied by  Seshadreesan and Ghosh in \cite{ghosh2011}. They have shown that
for a system of two spin-1 particles, the maximum of nonzero Hardy probability, i.e., $p_{\rm{max}}$ is the same
as for the case of two spin-1/2 particles, i.e., $p_{\rm{max}}=\frac{5\sqrt{5}-11}{2}$ and conjectured that 
the maximum remains the same for the system of two spin-$s$ particles. In an interesting development, this conjecture has been proved by Rabelo, Zhi and Scarani in \cite{scarani}. They have proved that $p_{\rm{max}}=\frac{5\sqrt{5}-11}{2}$   
irrespective of the dimension and the type of the system. Thus, $\frac{5\sqrt{5}-11}{2}$ can be thought of as an analogue of Tsirelson's bound \cite{cirelson} for Hardy's test of nonlocality.
In the following, we briefly outline the proof presented in \cite{scarani}.

The joint probabilities appearing in the nonlocality argument (\ref{minimalform}) takes the following form in quantum theory
\begin{equation}\label{proof1}
{\rm Prob}(A_k, a_k~;\ B_l, b_l)={\rm Tr}(\rho~\Pi_{a_k, A_k}\otimes \Pi_{b_l, B_l})
\end{equation}
where $\rho$ is the state of the system associated with an arbitrary Hilbert space $\mathcal{H}^{A}\otimes H^{B}$; $\Pi_{a_k, A_k}$ and   $\Pi_{b_l, B_l}$ are the measurement
operators corresponding to the outcomes $a_k$ and $b_l$ of measurements $A_k$ and
$B_l$ respectively. The measurements can be the most general  quantum measurement represented by the POVM elements. However, as here the  dimension of the Hilbert space is not constrained, hence measurements can be assumed to be projective by Neumark's theorem \cite{neumark} \cite{footnoteneumark}. The proof then uses the following lemma, proven in \cite{massanes1}:\\
{\bf Lemma 1:} Let $\Pi_{a_1, A_1}$, $\Pi_{\neg a_1, A_1}$,$\Pi_{a_2, A_2}$, $\Pi_{\neg a_2, A_2}$ be the four projectors acting on a Hilbert space $\mathcal{H}$ such that $\Pi_{a_1, A_1}+ \Pi_{\neg a_1, A_1}=\mathcal{I}$ and  $\Pi_{a_2, A_2}+ \Pi_{\neg a_2, A_2}=\mathcal{I}$ then there exists an orthonormal basis in $\mathcal{H}$ in which the four projectors are simultaneously block diagonal and the subspace $\mathcal{H}^i$ of $\mathcal{H}$ corresponding to the $i$th block is at most two dimensional.\\

The assumptions that $\Pi$'s appearing in Eq. (\ref{proof1}) are projectors and noticing that the above lemma is also valid in Bob's side, Eq.  (\ref{proof1}) can be rewritten as 
\begin{eqnarray}\label{diproof}
{\rm prob}(A_k, a~;\ B_l, b_l)&={\rm Tr}(\rho~\Pi_{a_k, A_k}\otimes \Pi_{b_l, B_l})\nonumber\\
&=\sum_{i,j}q_{ij}{\rm Tr}(\rho_{ij}~{\Pi^i}_{a_k, A_k}\otimes {\Pi^j}_{b_l, B_l})\nonumber\\
&\equiv \sum_{i,j}q_{ij}p_{ij}(A_k, a_k~;\ B_l, b_l)
\end{eqnarray}
where $q_{ij}={\rm Tr}(\rho~\Pi^i\otimes\Pi^j)$ and $\rho_{ij}=\frac{(\Pi^i\otimes\Pi^j\rho~\Pi^i\otimes\Pi^j)}{q_{ij}}$ is at most a two-qubit state; $\Pi^i$ and $\Pi^j$ denote the projectors onto $\mathcal{H}^{i}_{A}$ and $\mathcal{H}^{j}_{B}$ respectively. Since $q_{ij}\geq 0$ for all $i,j$ and $\sum_{ij}q_{ij}=1$, the first three conditions of (Eq.\ref{minimalform}) are satisfied iff they are also satisfied for each of the $p_{ij}$.  But then 
\begin{equation}\label{dimaxprob}
{\rm prob}(A_2,a_2~; B_2,b_2)=\sum_{ij} q_{ij} p_{ij}(A_2,a_2~; B_2,b_2)
\end{equation}
is a convex sum of nonzero Hardy's probabilities in each two-qubit subspaces. As a convex sum it is less than or equal to the largest element in the combination whose maximum value is known to be $\frac{5\sqrt{5}-11}{2}$ (section 4.4). This concludes the proof.

Recently, there has been  an yet another type of generalization of Hardy's conditions for bipartite higher dimensional system \cite{cabello2013} where probability of nonlocal events (maximum probability of success) grows with the dimension of the local systems and reaches approximately $40\%$ in the infinite limit. The fact that the success probability of this generalization of Hardy's argument increases with the dimension of local system has made it useful in witnessing the minimum Schmidt rank of an unknown state in a device-independent manner \cite{manikschmidt}, i.e., without knowing the details of the experimental devices involved in such experiments.  

\section{Hardy's nonlocality and generalized nonlocal theory}
Though quantum theory predicts nonlocal correlations, but this cannot be used to  
to communicate with a speed greater than that of the light in vacuum. 
But quantum theory is not the only non-local theory consistent with the relativistic
causality \cite{popescu941302.5296}. Theories which predict non-local correlations but are constrained with the no signalling condition are called ‘Generalized non-local theory (GNLT)’. In recent
years there has been considerable amount of interest in GNLT
\cite{masanes, scarani,pironio,jones, Barrett,cerf, Acin}. In
general, quantum theory has been studied in the background of
classical theory which is comparatively restrictive. The new idea
is to \emph{study quantum mechanics from outside},i.e.,
starting from this more general family of theories, and to study
properties common to all \cite{Acin}. This  might help in a better
understanding of quantum nonlocality. The Hardy’s non-locality argument 
in the framework of GNLT has been studied in \cite{choi,sk98}. It has been shown there that for a two-level system,
the maximum of Hardy's probability can be increased upto $0.5$ in a GNLT. Similar is the case with Bell's nonlocality argument. The maximum violation of Bell's inequality for bipartite systems in quantum mechanics is known to be $2\sqrt{2}$ (the Tsirelson's bound), but it can go up to $4$ without violating the relativistic causality \cite{popescu94}.    
This restricted nonlocal feature of quantum correlation has been explained by  the principle of Information Causality \cite{ic}  and also by the principle of Macroscopic Locality \cite{ml}. Each of these principles separately
explains the Tsirelson bound, but none of them explains the maximum success probability of Hardy's argument.
 Ahanj et al. have shown that the principle of Information causality can restrict the Hardy's probability
only up to 0.027 \cite{ahanj} whereas in case of macroscopic locality this value has been found to be  0.2062 \cite{manikml}. Thus,
reproducing the quantum value of Hardy's probability with the help of some physical principle(s) remains open.

\section{Hardy type nonlocality argument for multiparty}
The original logical structure of Hardy was extended to reveal the nonlocal 
feature of entangled states of three spin-1/2 particles \cite {ghosh98, wu}. Consider a situation where a physical system consisting of three two-level subsystems is distributed among three spatially separated parties, Alice, Bob and Charlie.
Each of these parties can randomly choose between the two $\pm1$-valued observables $U_j$ and $D_j$
(j=1,2,3 respectively for Alice, Bob and Charlie). A local-realistic reasoning cannot justify the simultaneous 
satisfaction of the following set of equations \cite {ghosh98, wu} 
\begin{equation}
\label{hardynon} \begin{array}{lcl} 
\rm{Prob} (D_1 = +1,U_2 = + 1, U_3 = + 1) &=& 0,\\
\rm{Prob} (U_1 = + 1,D_2 = + 1, U_3 = + 1 ) &=& 0,\\
\rm{Prob} ( U_1 = + 1,U_2 = + 1, D_3 = + 1 ) &=& 0,\\
\rm{Prob} (  D_1 = - 1,D_2 = - 1, D_3 = - 1 ) &=& 0,\\
\rm{Prob} (  U_1 = + 1, U_2 = + 1, U_3 = + 1
) &>& 0,
\end{array}
\end{equation}

However, there are entangled states of three qubit systems which satisfy these conditions \cite{wu, Kar(1997a),ghosh98} and thus exhibit nonlocality. Interestingly, in sharp contrast to bipartite cases, every maximally entangled state of three qubits \footnote{ any pure state of three qubits, which is locally unitarily
connected to the GHZ state $\frac{1}{\sqrt{2}}[|000\rangle+|111\rangle]$} exhibits Hardy's non-locality and for each of these states, probability of success of Hardy's argument can go maximum upto $12.5\%$ \cite{ghosh98}. In a subsequent development, $12.5\%$ was identified as the maximum success probability of this argument \footnote{maximized over every state and every set of observables} for three-qubit systems \cite{sk98}.
Reference \cite{sk98}  also studies this argument in the context of GNLT and finds
that the maximum probability reaches $50\%$ which is surprisingly the same as that for two
two-level systems. We conclude this paragraph by stating that an argument similar to that in section (4.3) will show that unlike the case of two-qubit systems, there can be mixed states of three-qubits which admit Hardy-type nonlocality \cite{Kar(1997a)}.

Can any pure entangled state of three qubits satisfy the Hardy-type conditions (\ref{hardynon})? 
This question was finally answered affirmatively in \cite {sk2010} succeeding the earlier attempts made in 
\cite {wu,ghosh98}. Now, from the set of joint probabilities in Hardy's or 
Hardy-like nonlocality without inequality
argument, one can, in principle, construct a linear inequality
involving these joint probabilities by using local realistic assumption.
\footnote{In the case of Hardy NLWI argument for two two-level
systems, this inequality (given in Eq. (11) of Ref. \cite{mermin94},
Eq. (11) of Ref. \cite{cabello2002}, and Eq. (26) of Ref. \cite{cereceda2004}) is nothing but
the corresponding CH inequality \cite{smroy}.}This inequality is automatically violated by every
quantum state which satisfies the corresponding NLWI argument.
This fact has been used in showing, that every three-qubit pure entangled
state violates a single linear inequality involving joint probabilities associated
with the Hardy-type NLWI (\ref{hardynon}). This is significant as for three-qubit systems, there is a family of entangled pure states, each of which satisfies all the Bell-type
inequalities involving correlation functions, arising out of
measurement of one between two noncommuting dichotomic
observables per qubit \cite{zukowski2002}. Though, later, Chen et al.\cite{chen2004} provided a
Bell-type inequality involving joint probabilities, associated to
measurement of one between two noncommuting dichotomic
observables per qubit, which is violated by all the states of the
above-mentioned family. But a single Bell-type inequality
was not guaranteed to be violated by all pure entangled states of three-qubits.

In \cite{cereceda2004}, the Hardy's argument has been extended to reveal the nonlocal features of $N$ spin-1/2 particles $N>3$. The author of \cite{cereceda2004} has shown that for all $N\geq 3$, any entangled GHZ state (including the maximally entangled ones) violates the Bell type inequality associated with the Hardy's conditions. This feature is again significant since it is known that  for all N odd, there are entangled GHZ states that do not violate any standard N-partite Bell inequality involving correlation functions \cite{zukowski2002}. 
Recently, as a major development in the field of multipartite nonlocality, by using Hardy type nonlocality without inequality argumentation, Yu et al.\cite{yu} have reported a single Bell-type inequality with a choice of two dichotomic observables per party which any pure entangled state of $N$-qubits violates. This result was then used to obtain
a Gisin-like theorem \footnote {Gisin's theorem states that all pure entangled state of two-qudit systems (two-$d$ dimensional quantum systems) violate a single Bell-type (CHSH) inequality with two measurement settings 
per sites and thus forbids local-realistic description for them \cite{gisin91}} for $N$-qudit systems by locally projecting a pure entangled state of such a system to the $N$-qubit subspace.

Though, the above generalizations of Hardy's nonlocality argument for N-qubit systems, witness entangled state but they do not differentiate between an entangled state and a genuinely entangled state of such systems.
By adding some more conditions in the above generalization, a new set of conditions has been provided in \cite{ramijzukowski} which can be resolved only by a genuinely entangled sate of $N$-qubits ($N\geq3$).
Moreover, for a given measurement settings, a unique pure genuinely entangled state of such systems 
($N$-qubit systems) satisfies these conditions. 
     
The structure of multipartite nonlocality is more complex than the bipartite case. Recent years have witnessed  
considerable amount of studies in the field of multipartite nonlocality. It all started when Svetlichny \cite{svetlichny} introduced the notion of genuine multipartite nonlocality. He also has provided a Bell-type inequality to detect genuine tripartite nonlocality of a system of three qubits. Later, this inequality was generalized to arbitrary number of parties \cite{collins}. This was further generalized to the most general scenarios   
involving an arbitrary number of parties and systems of arbitrary dimension \cite{bankal11}. However, Svetlichny's 
notion of genuine multipartite nonlocality suffers from a drawback. It does not exclude the probability distributions    
which can lead to signalling between bipartition. This leads to unphysical situations \cite{bankal11} and inconsistency from an operational point of view \cite{gallego}. In \cite{bankal11,almeida}, this fact has been taken into account to provide a new notion for genuine multipartite nonlocality. In a recent work, Chen et al. \cite{chen2014} have introduced a set of Hardy-like conditions keeping in view this new notion of genuine  nonlocality. A state satisfying these conditions exhibits genuine multipartite nonlocality. They have 
further shown that all pure entangled symmetric $n$-qubit ($n\geq3$) states are genuine multipartite nonlocal.     
In the case of asymmetric states, they have randomly chosen 50000 pure genuine multipartite entangled states for three and four qubit systems of and found all of them to satisfy the Hardy-like conditions. Thus they conjectured that all pure genuine multipartite entangled states are genuine multipartite nonlocal which remains to be proved .   

\section{Hardy's nonlocality and some  information  processing  tasks}

\subsection{ Hardy's nonlocality and true randomness }
We have seen previously that Hardy's conditions are not compatible with the assumption 
of local-realism. The assumption of local-realism is ontological in nature. In the following,
we show the incompatibility of Hardy's condition with an operational set of assumptions, namely the 
assumptions of predictability and no-signalling (also known as signal locality). 
The assumption of signal locality prevents one to send signals faster than light, predictability   
assumes that one can predict the outcomes of all possible measurements to be performed on the system.
In the context of typical Hardy's test described in section-2, we state below the two assumptions in precise mathematical forms:
\begin{itemize}
 \item A model is said to be predictable iff 
\begin{equation}\label{predictability}
 p(a,b|A, B, P)\in\{0,1\}~\forall~ a,b,A,B.
\end{equation}
\end{itemize}

\begin{itemize}
 \item A model is said to satisfy signal locality iff 
\begin{eqnarray}\label{nosig}
p(a|A, B, P)=p(a|A,P)~\forall~ a,A,B, \nonumber\\
p(b|A, B, P)=p(b|B,P)~\forall~ b,A,B.
\end{eqnarray}
\end{itemize}
As shown in \cite{wiseman} , the joint assumption of predictability and signal locality leads to the factorizability relation
of joint probability described in Eq. (\ref{factorizability}). The argument is as follows. As $ p(a,b|A, B, P)\in\{0,1\}$ hence conditioning it on further variable(s) cannot alter it, i.e.,
\begin{equation}\label{extraconditiong}
 p(a,b|A, B, P,\lambda)=p(a,b|A, B, P).
\end{equation}
Now according to Baye's theorem
\begin{equation}\label{bays1}
 p(a,b|A, B, P)=p(a|A,B,P)p(b|A,a,B,P).
\end{equation}
The assumption of predictability implies that $b= f(A,B,P)$ (i.e., $b$ is specified by specifying $A$, $B$, and $P$)
and hence 
\begin{equation}\label{predictability2}
 p(b|A,a,B,P)=p(b|A,B,P).
\end{equation}
Putting for $p(b|A,a,B,P)$ from Eq. (\ref{predictability2}) into Eq. (\ref{bays1}), we get
\begin{equation}
 p(a,b|A, B, P)=p(a|A,B,P)p(b|A,B,P)
\end{equation}
which from the assumption of signal locality can be rewritten as
\begin{equation}
p(a,b|A, B, P)=p(a|A,P)p(b|B,P) .
\end{equation}
Putting this into Eq. (\ref{extraconditiong}), we get
\begin{equation}\label{penunltiamte}
 p(a,b|A, B, P,\lambda)=p(a|A,P)p(b|B,P).
\end{equation}
By conditioning the RHS of the above equation on $\lambda$, we get the factorizability relation of Eq.(\ref{factorizability}).
The incompatibility of Hardy's conditions with the joint assumptions of predictability and no-siganlling
can  then be shown by arguing in the manner described in section-3. This incompatibility of Hardy's conditions with
assumptions of predictability and no-signalling thus imply that no predictable model can explain the satisfaction ofl the Hardy's
conditions if the model does not allow signalling. As no-signalling is in accordance with special relativity (and also experimentally testable in the present context), hence we conclude that true randomness is associated with the phenomena of satisfaction of Hardy's conditions. This fact plays the key role in  generation of true random numbers by using Hardy's argument. 

\subsection{Hardy's nonlocality and self testing of entangled states}
We have seen in section-4.4 that in case of two-qubit systems, the following specific state and measurement settings achieves the maximum probability of success for the Hardy's argument
\begin{eqnarray}\label{selftesting}
 | \psi \rangle = b(| u_1 \rangle \otimes | v_2 \rangle + |v_1 \rangle\otimes | u_2 \rangle)+e^{i\theta} \sqrt{1-2b^2} | v_1 \rangle \otimes | v_2\rangle,\nonumber\\
 A'= | u_1 \rangle \langle u_1 |-| v_1 \rangle \langle v_1 |;B'= | u_2 \rangle \langle u_2 |-| v_2 \rangle \langle v_2 |, \nonumber\\
A = | w_1^\perp \rangle \langle  w_1^\perp| -| w_1 \rangle \langle w_1 |;  B=| w_2^\perp \rangle \langle  w_2^\perp| -| w_2 \rangle \langle w_2 |,\nonumber\\
\end{eqnarray}
where
\begin{eqnarray}\label{selftestingbasis}
b=\sqrt{\frac{3-\sqrt{5}}{2}}, | w_1 \rangle = \frac{e^{i\theta} \sqrt{1-2b^2}| v_1 \rangle + b| u_1 \rangle} {\sqrt{1-b^2}},\nonumber\\ 
| w_2 \rangle = \frac{e^{i\theta} \sqrt{1-2b^2}| v_2 \rangle + b| u_2 \rangle} {\sqrt{1-b^2}};~
 {\rm{\theta~ is ~arbitrary ~and~ lies~ in~ [0,2\pi]}} .
\end{eqnarray}
In section -5.3, we have also seen that the maximum value of Hardy's probability for bipartite quantum systems is same as that for the two-qubits. The proof is also briefly presented there.
It follows from the proof that the nonzero Hardy's probability of Eq. (\ref{minimalform}) attains its maximum iff $p_{ij}(A_k, a_k~;\ B_l, b_l)$ (Eq.\ref{diproof}) is maximal for every $i,j$ such that $q_{ij}\neq 0$\cite{footnote2}.

These facts indicate that when the maximum Hardy's probability is observed, the state must somehow be a direct sum of copies of $\psi$ of Eq. (\ref{selftesting}) . In \cite{scarani} it has been shown that this indeed is the case. Thus, the Hardy's test constitute a self-testing of $\psi$.
Self-testing refers to the fact that some statistics predicted by quantum theory determine the state and the measurement upto a local isometry \cite{scaranireview,mayer}. This fact has been used in a number of device-independent information theoretic  tasks by using Hardy's correlations \cite{ramijrandom,ramijbyzantine, ramijcryptography}. In the following we briefly describe one such task.

\subsection{Device-independent generation of random numbers}  

Device independent generation of random numbers is very important from a practical point of view and so it has attracted much attention in recent years \cite{Pironio}. In device independent scenario, one does not have detailed knowledge about the experimental apparatuses and hence the experimental setup is like a black-box with inputs and outputs. At the input probes, one can change the parameters of measurement setup and thus can choose different measurements. The outputs are collected and the statistics is then analyzed to check whether the Hardy's conditions are satisfied. As we have seen above, that the probability distributions which satisfy all the Hardy's conditions cannot be explained by any predictable model and therefore true randomness is associated with it. The associated randomness is quantified by  by min-entropy \cite{Koenig} which is a statistical measure of the amount of randomness 
that a particular distribution contains. The maximal randomness can reach up to 1.35 if the corresponding Hardy's probability attains its maximum value $\frac{{5\sqrt{5}-11}}{2}$ \cite{ramijrandom}. This is interesting as the maximal min-entropy of 
the generated randomness is about 1.23 if the output probabilities are analyzed to see the violation of a CHSH-inequality instead. This maximum occurs when the CHSH expression reaches its maximum (the Tsirelson bound).

  \section{Hardy's argument for temporal nonlocality}
For testing the existence of superposition of macroscopically distinct
quantum states, Leggett and Garg \cite{legett} put forward the notion of {\it macrorealism}.
This notion rests on the classical paradigm \cite{brukner1, usha} that (i) physical properties of a
macroscopic object exist independent of the act of observation and (ii) it is possible, at least in principle, to determine these properties, without
any effect on the state itself or on its subsequent dynamics.

These original assumptions of \cite{legett}, namely the assumptions of  `macroscopic realism' and 
`noninvasive measurability', have been generalized to derive a temporal version of the Bell-CHSH  
inequality irrespective of whether the system under consideration is macroscopic or not 
\cite{brukner, generalized}.
Unlike the original Bell-CHSH scenario \cite{chsh} where correlations between measurement results from two distantly located physical systems are considered, temporal Bell-CHSH inequalities (or its generalizations)
are derived by focusing on one and the same physical system and analyzing the correlations
between measurement outcomes at two different times. These derivations are based on the following two assumptions: 
(i) {\it Realism}: The measurement results are determined by (possibly hidden) properties, which the particles
carry prior to and independent of observation, and (ii) {\it Locality in time}: The result 
of a measurement performed at time $t_2$ is independent of any  ideal measurement performed at some earlier
or later time $t_1$ . 
   
These inequalities get violated in quantum mechanics and thereby give rise to the notion of
{\it entanglement in time} which has been a topic of current research interest 
\cite {brukner,usha,ali,fritz,joag,white,brukner2,reznik1,reznik2,home}. These inequalities have been studied to probe
the similarities and differences between spatial and temporal correlations in quantum mechanics
to learn more about the relation between the structure of space and time and the abstract formalism
of quantum theory \cite{brukner}. Apart from the theoretical significance, this study 
also has practical implications. The correspondence between the communication costs of 
the classical simulation of spatial correlations 
and the memory costs in the simulation of temporal correlations can be taken as an example \cite{brukner}. The \emph {entanglement in time}, has been shown to save the size of classical memory required in certain computational problems beyond the classical limits.

Interestingly, the original argument of Hardy, which establishes the incompatibility of 
quantum theory with the notion of local-realism, can also be used to reveal this time-nonlocal feature of quantum states \cite{ali,fritz, white}. To see this we consider a single two- level physical system on 
which an observer (Alice) chooses to measure
one of two observables $A$ or $A^{'}$ at time $t_1$, whereas at a later time $t_2$,
another observer (Bob) \cite{bob} measures either of the two observables  $B$ and $B^{'}$. 
Consider now the  set of conditions (\ref{hardy2q1})-(\ref{hardy2q4}). 
The condition (\ref{hardy2q4}) says that if Alice chooses to measure the observable $A^{'}$ and 
Bob  chooses observable $B^{'}$, he will not obtain  $+1$ as measurement result
whenever Alice has  detected the measurement value $+1$. The remaining conditions can be analyzed in a similar manner. This version of Hardy's argument makes use of the fact
that not all of the conditions (\ref{hardy2q1})-(\ref{hardy2q4}) can be simultaneously satisfied in a time-local realistic theory, but they can be in quantum mechanics.  

In a realistic theory, values are assigned to all the observables 
(whether or not they are actually measured) in such a manner that they 
agree with experimental observations. Consider a situation where a realist has been 
supplied with a table asking for the values of 
$A$, $A^{'}$, $B$ and $B^{'}$ in several runs of a Hardy experiment. 
In order to satisfy the first condition, he will have to assign $+1$ for $A$  and $+1$ for $B$ a few times. 
Out of these few times,  he cannot choose the value  $-1$ for observable $B^{'}$ since this event has zero probability according to the condition (\ref{hardy2q3}). Now the second condition implies  either $\rm {Prob}(-1|A^{'})=0$ or  $\rm{Prob}(+1|B)=0$ whenever $A^{'}$ is $-1$.
Thus the realist,  will either always assign $+1$ for $A^{'}$ or he will have to assign $-1$ for $B$ whenever he assigns $-1$ for $A^{'}$ . He is free to assign
$+1$ or $-1$ for $A^{'}$, but the columns of the table where he has put $+1$ for $B$, he cannot put $-1$ for $A^{'}$. This is because
if he puts $-1$ for $A^{'}$, he will have to put $-1$ for $ B$ which is not possible according to the assumption of {\it nonlocality in time}.  Thus his table will have columns  with a few entries with $+1$ for all the four observables $A, A^{'}, B$ and $B^{'}$ and thus the first condition (\ref {hardy2q1}) gets violated. However, all these conditions can be satisfied in quantum theory. As an example, we write below a state and the observables which satisfy all the conditions;
\begin{equation}\nonumber
 |\psi\rangle=\left( \begin{array}{cccc}
0 \\
 1  \end{array} \right) 
\end{equation}

\begin{equation}\nonumber
A=B^{'}=\left(\begin{array}{ccc}
0 &1 \\
1 & 0 \end{array} \right)
\end{equation}
\begin{equation}\nonumber
A^{'}=B=\left(\begin{array}{ccc}
1 &0 \\
0 &-1 \end{array} \right)
\end{equation}
It can also be checked that for the above set, the nonzero probability (\ref{hardy2q1}) is $25\%$ which is also the maximum success probability of this argument for qubit systems \cite{ali,fritz}. The experimental verification of the above fact followed soon after in \cite{white}. Recently, the temporal version of Hardy's argument has been studied for any finite dimensional systems \cite{choudhary}.  Ref \cite{choudhary} shows the maximum of the nonzero probability appearing in the temporal version of Hardy's argument (\ref{minimalform})  remains $25\%$ irrespective of the dimension of the quantum mechanical system and the type of observables involved.  For the special case of spin measurements, for spin-1 and spin-3/2 observables, the authors of \cite{choudhary} have shown that this maximum can also be achieved. They have further conjectured that this maximum  can be observed for any spin system. 

\section{conclusion}
In conclusion, we have reviewed the Hardy's argument and its various generalizations which reveals the nonlocal features of quantum states. We have also discussed a few information theoretic tasks which make use of this argument for their successful completions.

\section{Acknowledgment:} 
 S.K.C. acknowledges support from the Council of Scientific and
Industrial Research, Government of India (Scientists' Pool Scheme).

\section*{References}


\begin{thebibliography}{9}
\bibitem{footnoteneumark} By Neumark's theorem the probability distributions of the outcomes of POVMs can be obtained  by projective measurements on systems in higher dimensional Hilbert space.
\bibitem{footnote2} It should also be noted that for the maximum value to be reached it is necessary that for all $i$ an $j$ such that $q_{ij}\neq 0$, the dimension of both $\mathcal{H}^i$ and $\mathcal{H}^j$ be equal to two.
\bibitem{generalized} There is one more difference between \cite{legett} and \cite{brukner}: while in \cite{legett}, the
observer measures a single observable at different times under the supposition that its value is 
independent whether or not a measurement (of the same observable) was performed at some earlier or later time,
in \cite{brukner} at any given time the observer has a choice between two (or more) different measurement
settings. The value assigned to an observable $A$ at time $t_1$ is independent of
whether or not a measurement was performed at some different time $t_0$ or
which observable at that time was measured.
\bibitem{bob} Alice and Bob may be one and the same observer, but the randomness in the choice of $A_1$ 
or $A_2$ and that of $B_1$ or $B_2$ must be independent.
\end{thebibliography}
\end{document}